\documentclass[apj,onecolumn,onecolappendix,numberedappendix,appendixfloats,iop]{openjournal}

\usepackage{booktabs}
\usepackage{longtable}

\setcitestyle{numbers}
\setcitestyle{square}

\usepackage{color}
\usepackage{xcolor, colortbl}
\usepackage{comment}
\usepackage{amsmath, amssymb, graphics, setspace}
\usepackage{multirow}
\usepackage{longtable}
\usepackage{tabularx,ragged2e}
\newcolumntype{L}{>{\RaggedRight}X}

\usepackage{tikz}

\begin{document}

\title{Map Multi-Tool: A Map-Based Approach to Modeling Beam Systematics for Cosmic Microwave Background Experiments}
\author{King,~C.L.$^{1,*}$}
\author{Hryciuk,~A.$^{2}$}
\author{McMahon,~J.$^{2}$}
\author{Nagy,~J.M.$^{1}$}
\author{Ruhl,~J.E.$^{1}$}
\affiliation{$^{1}$Case Western Reserve University, Physics Department}
\affiliation{$^{2}$The University of Chicago Physics Department}
\email[$^*$Corresponding Author: ]{cxk554@case.edu}

\begin{abstract}
Cosmic microwave background (CMB) experiments use simulations of instrumental systematic effects to ensure high-fidelity measurements of cosmological parameters.  Quantifying the expected magnitude of these effects enables experiments to improve designs, set performance requirements, and understand potential measurement biases from residual systematics. Here we present a new simulation framework, called Map Multi-Tool (MMT), which models beam-related systematics for CMB instruments using a map-based approach. The pipeline convolves simulated sky realizations with distorted intensity and polarization beams, including leakage effects, to produce sky maps and power spectra. These outputs can then be used as inputs for cosmological parameter estimators.  This framework enables efficient evaluation of such 
systematic effects.  We demonstrate the capabilities of MMT with examples of non-ideal beams induced by electrical readout crosstalk and detector time constant response.  The electrical crosstalk example considers eight different schemes for a time-division multiplexed readout architecture, and shows how they lead to different levels of angular power spectrum leakage for row-switching and inductive crosstalk. The detector time constant example demonstrates how associated uncertainties can alter CMB spectra at high multipoles and bias derived cosmological parameters. These examples illustrate some of MMT’s broad capabilities to inform critical design and calibration decisions that mitigate systematic effects in CMB instruments.
\end{abstract}

\maketitle

\section{Introduction}
Current and upcoming Cosmic Microwave Background (CMB) experiments are targeting a variety of cosmological and astrophysical science goals.  These include searching for the primordial gravitational waves predicted by cosmic inflation and providing more stringent constraints on the thermal history of the universe that could detect or rule out particles beyond the Standard Model \cite{S4_science_book, SO_science_book}. Achieving these goals requires high-precision measurements with rigorous control of instrumental systematics, particularly those affecting the telescope beam response. Beam-related systematics can cause distortions that may induce signal leakage between temperature ($T$) and $E$- and $B$-mode polarization anisotropies, which are given by rotationally invariant linear combinations of Stokes $Q$ and $U$ \cite{EB_conversion}. Such distortions could impact cosmological constraints inferred from these anisotropies, including those related to inflation (the tensor-to-scalar ratio $r$) and the effective number of relativistic species ($N_{\mathit{eff}}$). Managing these effects through instrument design, calibration, and data analysis requires a quantitative understanding of their behavior. This can be facilitated by an efficient simulation pipeline that captures anticipated systematics and propagates them through a realistic analysis procedure. 

Advances in large-format detector arrays \cite{DRM_SPIE, horn_dets, BKS_dets} and novel telescope designs \cite{BK18_2021,act,spt3g,SO} have enabled the deployment of increasingly sensitive instruments with tens of thousands of detectors. Detector counts will continue to grow in next-generation experiments \cite{S4_ref_design, SO_upgrade}, creating new challenges for realistic end-to-end simulations. Previous efforts range from analytic calculations (\emph{e.g.}, \cite{Hu_Hedman_Zaldarriaga, shimon_keating}) to full time-ordered-data (TOD) simulations (\emph{e.g.}, \cite{Planck_tod_sims, TOAST}). Each of these methods has limitations. Analytical calculations may be unable to capture the full complexity of the signal leakage. TOD simulations offer a powerful alternative, but they are  computationally expensive, which can limit their scope or feasibility in some design studies.

In this paper, we present a complementary, computationally-efficient modeling framework called Map Multi-Tool (MMT).
MMT generates realizations of instrument-beam systematics directly in the map domain, enabling analysis in the power-spectrum domain and subsequent propagation to cosmological-parameter estimation. This tool can be used to model the impact of a wide variety of beam-related systematics on the resulting measured angular power spectra. Furthermore, it allows for the evaluation of individual systematic effects as well as the combined interactions of multiple systematics to model complex instruments.

Here we describe the computational approach implemented in MMT and illustrate its capabilities with two examples of beam-related systematics. Section \ref{sec:formal} outlines the mathematical formalism of MMT; Section \ref{sec:examples} applies it to electrical crosstalk and detector time constant effects; Section \ref{sec:conclusions} concludes with a discussion of these results and implications for future work.  

\section{Formalism}
\label{sec:formal}

CMB instruments typically produce sky maps of temperature and linear polarization using the Stokes parameters $I$, $Q$, and $U$. Although the circular polarization component, Stokes $V$, can also be measured in principle, both theoretical predictions and experimental measurements suggest that it is much smaller than the linear polarization, so it is typically neglected \cite{Class_Vpol, Spider_Vpol}.

The MMT framework models the maps observed by an instrument as
\begin{equation}
    S_{obs}=M \circledast S_{sky} + \mathcal{N}.
    \label{eqn:MMT_model}
\end{equation} 
Here $\circledast$ represents a convolution, $S_{obs}$ is a Stokes-parameter vector of the observed maps, $M$ is a $3\times3$ matrix of maps of Mueller matrix elements, $S_{sky}$ is the input Stokes-parameter vector of maps generated from a simulated CMB sky with known cosmological parameters, and $\mathcal{N}$ is a Stokes vector of instrument noise maps. Each of these noise maps can include both white and non-white components from sources such as the atmosphere or ground. Explicitly writing each element in $S_{obs}$, $M$, $S_{sky}$, and $N$, Equation \ref{eqn:MMT_model} becomes:
\begin{equation}
\begin{pmatrix}I_{obs} \\ Q_{obs} \\ U_{obs}\end{pmatrix} 
= 
\begin{pmatrix}M_{II} & M_{IQ} & M_{IU} \\ 
               M_{QI} & M_{QQ} & M_{QU}  \\ 
               M_{UI} & M_{UQ} & M_{UU} \\ 
            
               \end{pmatrix}
\circledast
\begin{pmatrix}I_{sky} \\ Q_{sky} \\ U_{sky}\end{pmatrix} 
+
\begin{pmatrix}\mathcal{N}_{I} \\ \mathcal{N}_{Q} \\ \mathcal{N}_{U}\end{pmatrix}. 
\label{eqn:MMT_model_expanded}
\end{equation}
Any systematic effect that distorts the instrument beam or induces coupling between $I$, $Q$, and $U$ therefore modifies the observed Stokes maps as described by Equation \ref{eqn:MMT_model_expanded}.

\begin{figure}[b]
\centering
\includegraphics[width=0.65\linewidth]{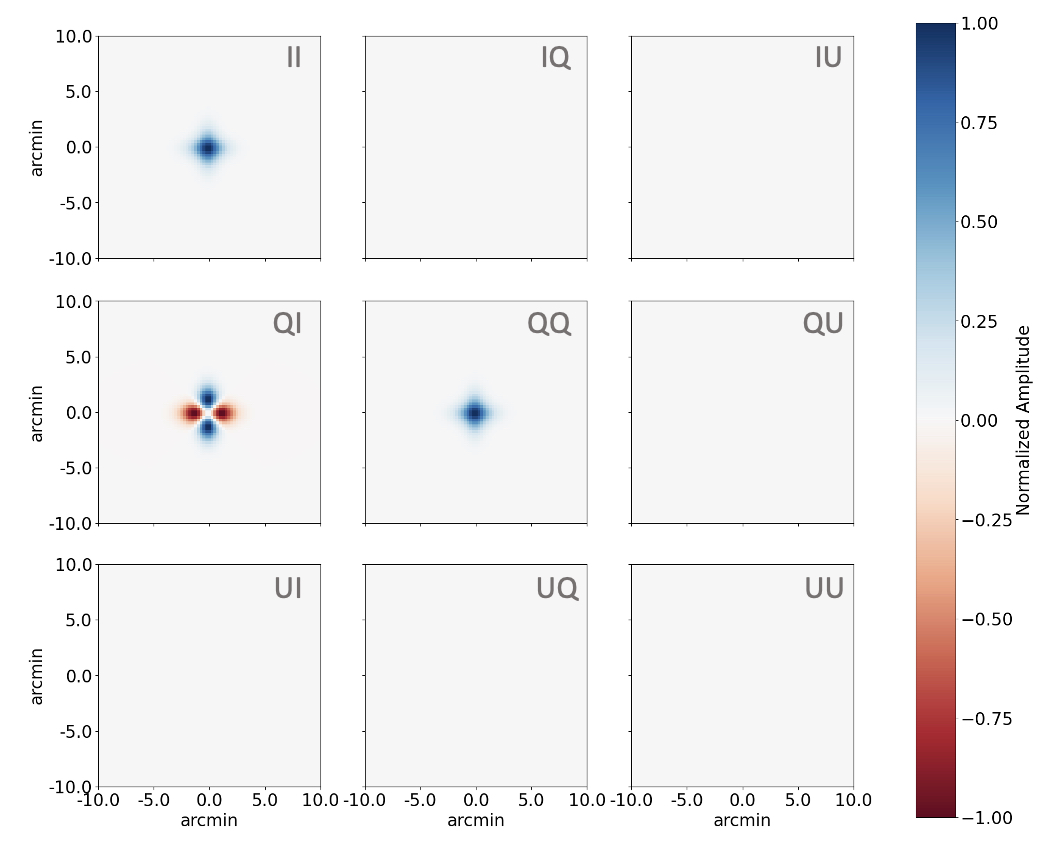}
\caption{The beam mixing matrix $M$ for $I \rightarrow (Q,U)$ coupling due to quadrupole beam mismatch for a pair of orthogonal $Q$-sensitive detectors. The diagonal elements illustrate the instrument beams for a given Stokes parameter while the off-diagonals show the coupling between them. In this example, the quadrupole orientation couples $I$ only to $Q$, but other orientation angles would also couple to $U$}.
\label{fig:beam_mismatch}
\end{figure}

To illustrate the beam mixing matrix $M$, we consider a case with non-zero $M_{QI}$ and $M_{UI}$ terms, coupling $I_{sky}$ into $Q_{obs}$ and $U_{obs}$. To connect this example to real instrumentation, we note that CMB polarization signals are commonly measured by differencing orthogonally-oriented, physically co-located detector pairs (\emph{e.g.}, \cite{bicep2_pairdiff,polarbear_pairdiff}). While this allows rejection of the common-mode temperature signal, it can introduce a systematic bias if the beam shapes are slightly different between the paired detectors. Figure \ref{fig:beam_mismatch} illustrates the post-differencing beam residual for the case where the beam mismatch for a set of pair-differenced detectors is exactly a quadrupole.  Beam mismatches with different, more complex geometries can be modeled in a similar way using a perturbative expansion (\emph{i.e.~}monopole, dipole, quadrupole, etc.)~to capture effects such as differential gain, differential pointing, and differential beam width and ellipticity \cite{bicep2_deproj}. Similarly, using only $E$-mode input signals allows for the analogous calculation of 
$E$-to-$B$ leakage.

To understand how a given systematic affects cosmological parameter inference, MMT transforms the simulated maps into the power spectrum domain under a flat-sky approximation by taking the Fourier transform of $S_{obs}$, denoted by $\tilde{S}_{obs}$. Throughout this work, we use the tilde to signify Fourier space, where there flat sky $|k|$ is mapped to $\ell$ for display of angular power spectra. The quantities $\tilde{Q}_{obs}$ and $\tilde{U}_{obs}$ are then converted into rotationally invariant $E$- and $B$-mode components, $\tilde{E}_{obs}$ and $\tilde{B}_{obs}$, following the formalism of \cite{EB_conversion}. The resulting power spectra can then be analyzed using standard cosmological parameter pipelines. These include simple Markov-Chain Monte-Carlo (MCMC) or maximum likelihood approaches as well as more complex parameter fitting packages such as COBAYA \cite{Cobaya}. This enables a quantitative assessment of how a particular beam systematic, encoded in the Mueller matrix $M$, biases constraints on cosmological parameters. 

MMT also computes the signal leakage between the observed spectra, which is another metric for evaluating the impact of beam systematics. When $S_{sky}$ contains only one component (\emph{e.g.}, pure intensity $I_{sky}$ or pure $E$-mode polarization), the leakage can be isolated through ratios of observed input power spectra. For instance, if $S_{sky}$ includes only $I_{sky}$, the fractional $T$-to-$B$ leakage is given by the ratio $\tilde{B}_{obs}/\tilde{I}_{sky}$.  

\section{Examples}
\label{sec:examples}

\subsection{Electrical Detector Readout Crosstalk}

We first illustrate the power of the MMT framework by considering instrumental systematics due to electrical detector readout crosstalk. Modern CMB experiments employ multiplexed readout architectures to efficiently handle their large detector counts. However, this introduces electrical crosstalk, in which signals from one detector couple to others and introduce correlations that distort the effective instrument beam. The particular character of the beam distortion depends on the spatial distribution of the correlated detectors and the wiring scheme. 

\begin{figure}[b]
\centering
\includegraphics[width=0.68\linewidth]{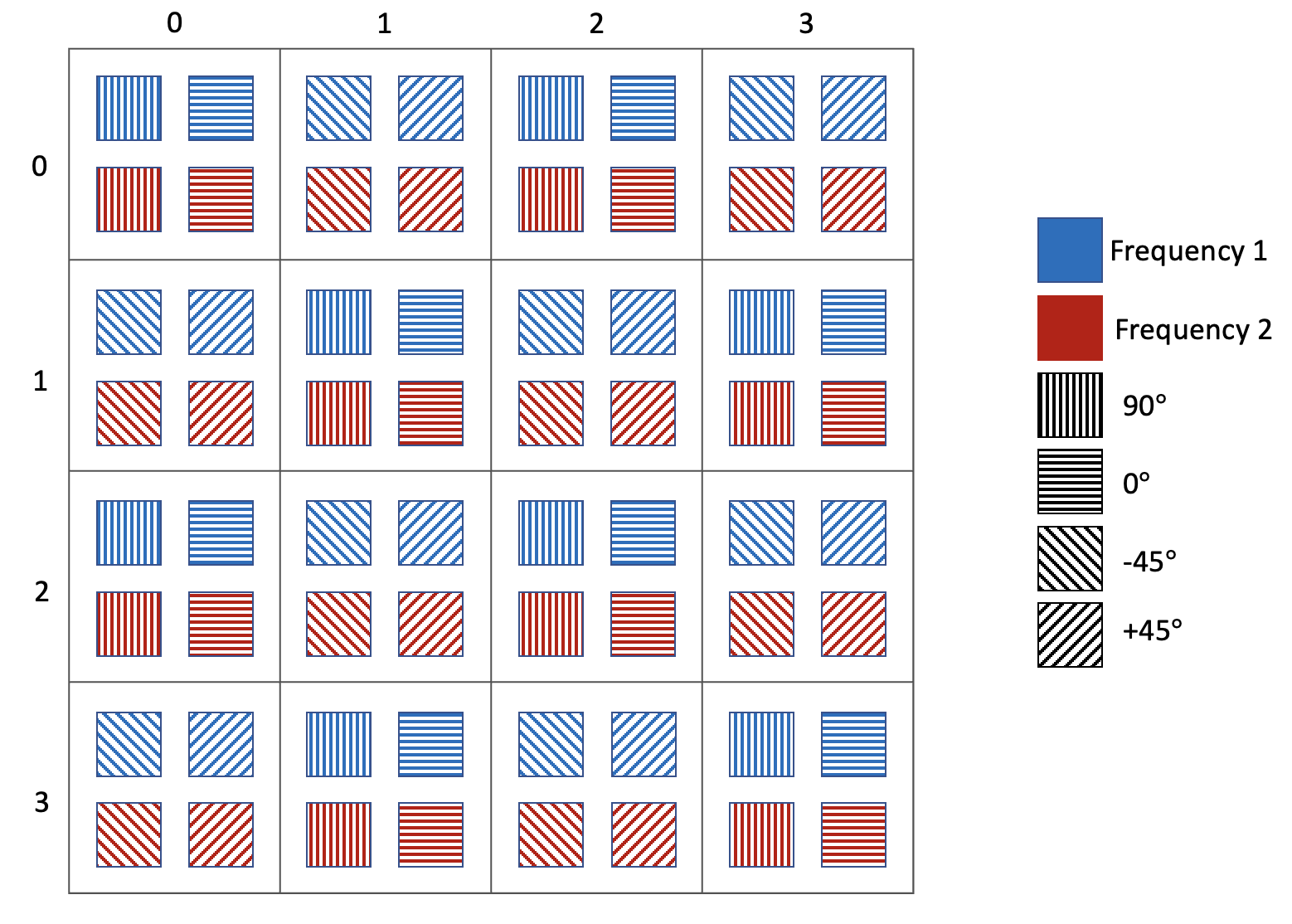}
\caption{Schematic diagram of the simplified detector array model used for electrical crosstalk simulations. Each spatial pixel in the $4\times4$ grid contains four individual detectors, which measure two different linear polarization orientations in two frequency bands. The relative polarization orientations of neighboring pixels are rotated by 45$^\circ$ with respect to each other, which provides better instantaneous coverage of both the $Q$ and $U$ Stokes parameters when all detector signals are combined.}
\label{fig:toy_model}
\end{figure}

In the following example, we consider a simplified detector array model observing the sky through a large aperture telescope, with pixel separations and beam sizes appropriate for a 5-meter, f/2.8 large aperture telescope similar to \cite{padin2018three, Gallardo24}. The array consists of a $4\times4$ grid of dichroic dual-polarization spatial pixels, as shown in Figure \ref{fig:toy_model}, with a 5.3~mm pixel spacing. Each spatial pixel contains four detectors, two measuring light in each of two different frequency bands,  with instrument beam sizes of 2.5' and 1.5' in the lower (90~GHz) and higher (150~GHz) bands, respectively. Each like-frequency detector pair on a pixel is sensitive to orthogonal linear polarizations. Half of the pixels contain detectors sensitive to 0$^\circ$ and 90$^\circ$, while the other half contain detectors sensitive to 45$^\circ$ and -45$^\circ$. The pixels are arranged in a checkerboard pattern, such that neighboring pixels are sensitive to different orthogonal polarization pairs.

We model the detectors as being connected to a time-division multiplexing (TDM) readout system (e.g.~\cite{DRM_SPIE}, \cite{S4_TDM}, \cite{TDM_2level}). In TDM architectures, detectors are arranged in a logical row–column grid, where all detectors in a single column share common voltage bias and readout lines. Individual detectors are selected by activating one row at a time using flux-controlled switches. The current from each detector is inductively coupled to a dedicated SQUID amplifier, and each column of SQUID amplifiers is connected to additional readout electronics. 

We consider two ways in which electrical crosstalk can arise in this system. The first is caused by the inductive coupling between neighboring SQUIDs, introducing correlations between detectors that are physical neighbors. We refer to this type of crosstalk as inductive crosstalk. The second occurs if the rate of TDM row switching is not sufficiently fast to accommodate the settling time of the readout, which introduces correlations between the detector currently being read out and the previous one. We refer to this type of crosstalk as row-switching crosstalk.  In this example, we model inductive crosstalk with $\sim$0.3\% correlations and row-switching crosstalk with $\sim$0.1\% correlations \cite{TDM_2level}, which are roughly consistent with typical levels for modern TDM systems.

MMT models detector correlations from both these types of electrical crosstalk as beam distortions. Mathematically, we consider the signal coupled from the $j$th detector (pointed at sky pixel $m$) into the $i$th detector (pointed at sky pixel ($n$), and then sum over all such couplings while accounting for the relative pointing of those two detectors. Ignoring circular polarization and setting all gains to unity, the sky-only signal in the $i$th and $j$th detectors are 
\begin{equation}
     d_i = I_n + \cos{2\theta_i}Q_n + \sin{2\theta_i}U_n 
\end{equation}
\begin{equation}
      d_j = I_m + \cos{2\theta_j}Q_m + \sin{2\theta_j}U_m, 
\end{equation}
where $\theta_i$ and $\theta_j$ characterize the detectors' linear polarization angles on the sky. With crosstalk, a portion $\gamma_{ij}$ of $d_j$ is coupled into the signal of detector $i$. Thus, the full signal from detector $i$ is given by
\begin{equation}
    \tilde{d_i}=d_i + \gamma_{ij}d_j.
\end{equation}
The mapmaking equation tells us how the 
signal from detector $i$ contributes to the intensity and polarization signal reconstruction at sky pixel $n$,  
\begin{equation}
\begin{pmatrix}\tilde{I_n} \\ \tilde{Q_n}\\ \tilde{U_n} \end{pmatrix} 
= 
\begin{pmatrix}1 & 0 & 0 \\ 
               0 & 2 & 0  \\ 
               0 & 0 & 2                
\end{pmatrix}
\begin{pmatrix} \tilde{d_i} \\ 
    \tilde{d_i} \cos{2 \theta_i} \\ 
    \tilde{d_i} \sin{2 \theta_i} 
\end{pmatrix} ,
\label{eqn:ij_coupling_matrix_eqn}
\end{equation}
as described in \cite{mapmaking_better}. Using this formalism we can find all the couplings of (I,Q,U) at other map pixels into the reconstructed map values at pixel $n$, caused by the couplings of other detectors into detector $i$. We construct a 3$\times$3 matrix containing nine N$\times$N pixel maps of the full signal response of detector $i$ ($\tilde{d_i}$) for each possible (I,Q,U) coupling:
\begin{equation}
    m_i=\begin{pmatrix}m_i^{II} & m_i^{IQ} & m_i^{IU} \\ 
               m_i^{QI} & m_i^{QQ} & m_i^{QU}  \\ 
               m_i^{UI} & m_i^{UQ} & m_i^{UU} \\ 
               \end{pmatrix}.
\end{equation}
The expected signal of detector $i$ without crosstalk ($d_i$) is stored in the center pixel of each map:
\begin{equation}
\begin{aligned}
    m_i^{II}[N/2,N/2] &= 1 & m_i^{IQ}[N/2,N/2] &= \cos{2\theta_i}  & m_i^{IU}[N/2,N/2] &= \sin{2\theta_i}\\
    m_i^{QI}[N/2,N/2] &= 2\cos{2\theta_i} & m_i^{QQ}[N/2,N/2] &= 2\cos^2{2\theta_i} & m_i^{QU}[N/2,N/2] &= 2\sin{2\theta_i}\cos{2\theta_i}\\
    m_i^{UI}[N/2,N/2] &= 2\sin{2\theta_i} & m_i^{UQ}[N/2,N/2] &= 2\sin{2\theta_i}\cos{2\theta_i} & m_i^{UU}[N/2,N/2] &= 2\sin^2{2\theta_i}.\\
\end{aligned}
\end{equation}
The coupling signal from detector $j$ appears in the pixel offset from the center pixel according to the physical offset between detectors $i$ and $j$. This offset is given by $\delta_{ij}^x$ in the x-direction and $\delta_{ij}^y$ in the y-direction. Thus, the signal from detector $j$ that couples into detector $i$ appears in map pixels located at $[x,y]$, where $x=N/2+\delta_{ij}^x$ and $y=N/2+\delta_{ij}^y$:
\begin{equation}
\begin{aligned}
m_i^{II}[x,y] &= \gamma_{ij}  &  m_i^{IQ}[x,y] &= \gamma_{ij} \cos{2\theta_j}  &  m_i^{IU}[x,y] &= \gamma_{ij} \sin{2\theta_j}\\
m_i^{QI}[x,y] &= 2\gamma_{ij}\cos{2\theta_i}  &  m_i^{QQ}[x,y] &= 2 \gamma_{ij} \cos{2\theta_i}\cos{2\theta_j}  &  m_i^{QU}[x,y] &= 2\gamma_{ij} \cos{2\theta_i}\sin{2\theta_j}\\
m_i^{UI}[x,y] &= 2\gamma_{ij}\sin{2\theta_i}  &  m_i^{UQ}[x,y] &= 2 \gamma_{ij} \sin{2\theta_i}\cos{2\theta_j}  &  m_i^{UU}[x,y] &= 2\gamma_{ij} \sin{2\theta_i}\sin{2\theta_j}.\\
\end{aligned}
\end{equation}
This is repeated for all detectors coupling into detector $i$. Analogous maps are constructed for each of the $n_{det}$ detectors and averaged to construct a matrix of maps $m$ representing the signal response across all detectors. This matrix is convolved with the instrument beam response $B$ to construct the mixing matrix $M$ introduced in the formalism described in Section \ref{sec:formal}:
\begin{equation}
    M = \left ( \frac{1}{n_{det}}\sum_i m_i \right ) \circledast B.
\end{equation}

Continuing to follow the formalism described in Section \ref{sec:formal}, MMT quantifies the resulting spectral leakage produced by crosstalk. Such leakage distorts and cross-couples the 
CMB power spectra and can subsequently bias constraints on cosmological parameters. The very weak, and still undetected, 
$B$-mode signal from cosmic inflation is especially susceptible to such effects, making it important to minimize such $T$-to-$B$ and $E$-to-$B$ leakage.

\begin{figure}[h]
    \centering
    \includegraphics[width=0.68\linewidth]{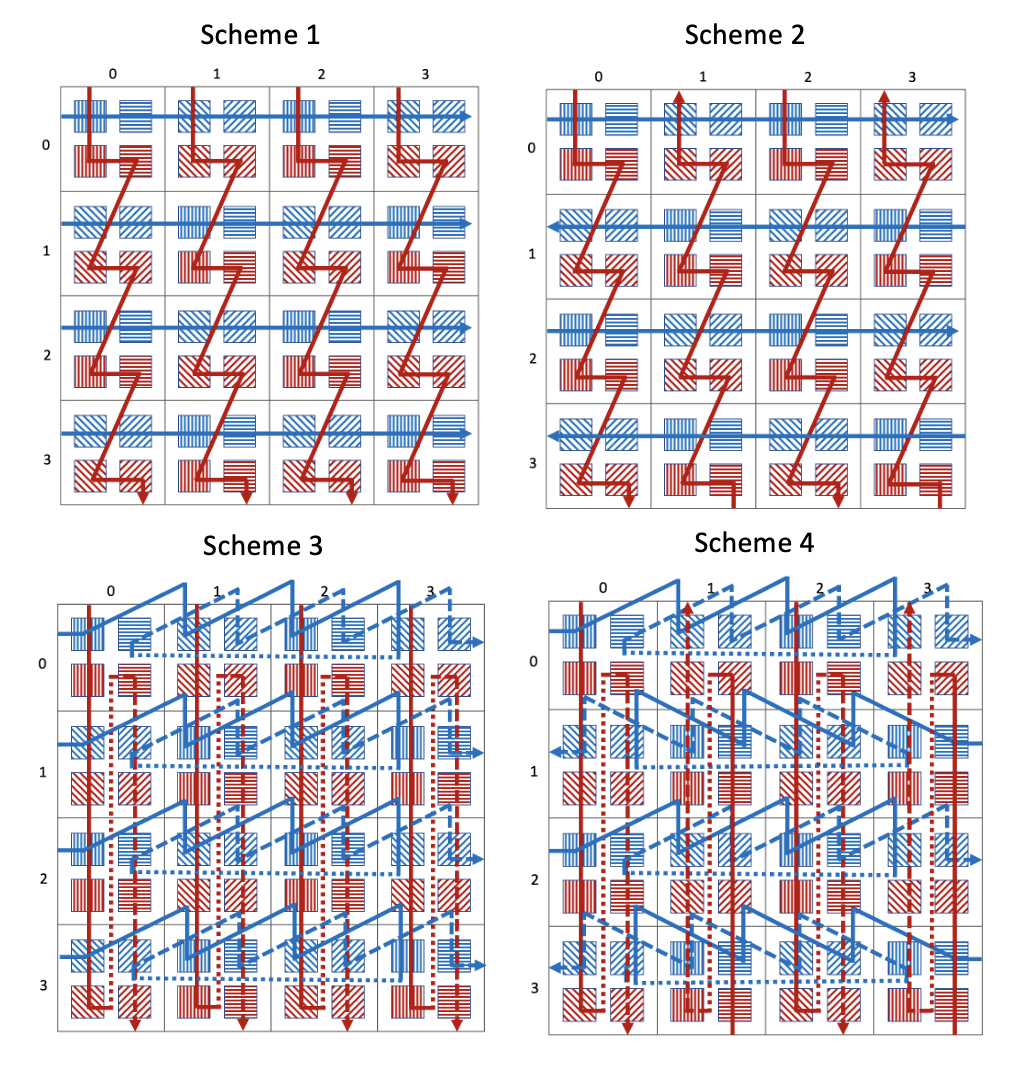}
    \caption{Diagrams for each of the four single-frequency readout schemes considered for the detector array shown in Figure \ref{fig:toy_model}. Blue and red arrows indicate the order in which Frequency 1 and Frequency 2 detectors are read out each row and column. Solid arrows represent the first pass through a row or column, while dashed arrows indicate a second pass through the row or column. \textit{Scheme 1} consecutively reads out both polarizations in a pixel before moving on to the next pixel. For Frequency 1, pixels are read out from left to right per row. For Frequency 2, pixels are read out from top to bottom per column. \textit{Scheme 2} consecutively reads out both polarizations in a pixel before moving on to the next pixel. For Frequency 1, pixels are read out from left to right in even rows and right to left in odd rows. For Frequency 2, pixels are read out from top to bottom in even columns and bottom to top in odd columns. \textit{Scheme 3} consecutively reads out one polarization per pixel on all detectors before reading out the remaining polarization. For Frequency 1, pixels are read out from left to right per row. For Frequency 2, pixels are read out from top to bottom per column. \textit{Scheme 4} consecutively reads out one polarization per pixel on all detectors before reading out the remaining polarization. For Frequency 1, pixels are read out from left to right in even rows and right to left in odd rows. For Frequency 2, pixels are read out from top to bottom in even columns and bottom to top in odd columns.
}
    \label{fig:single_schemes}
\end{figure}

\begin{figure}[h]
    \centering
    \includegraphics[width=0.68\linewidth]{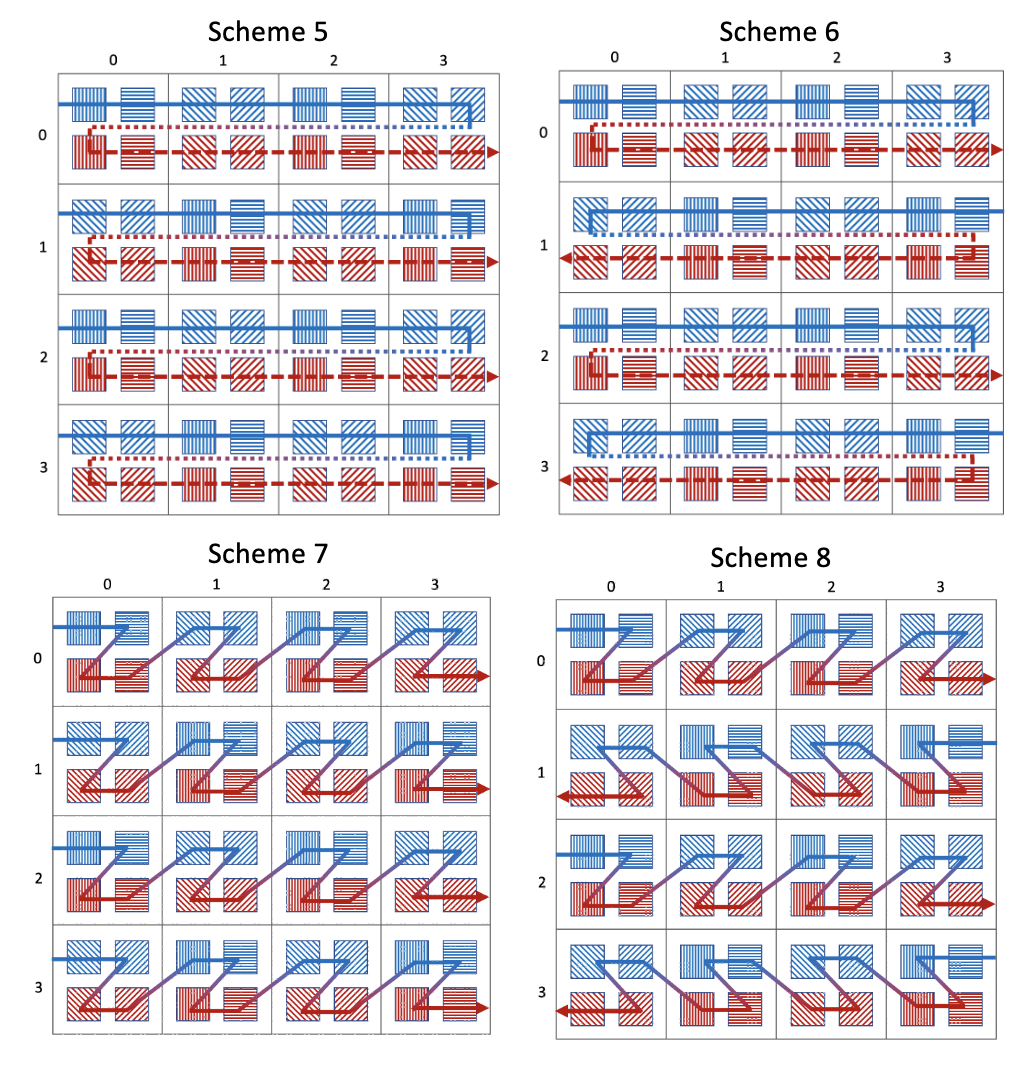}
    \caption{Diagrams for each of the four dual-frequency readout schemes considered for the detector array shown in Figure \ref{fig:toy_model}. Blue and red arrows indicate the order in which Frequency 1 and Frequency 2 detectors are read out across each row. Solid arrows represent the first pass through a row, while dashed arrows indicate a second pass through the row. \textit{Scheme 5} consecutively reads out both polarizations of Frequency 1 detectors in each pixel, moving left to right through each row. After all Frequency 1 detectors in a given row have been read out, consecutively read out both polarizations of Frequency 2 detectors in each pixel, moving left to right through the row. \textit{Scheme 6} consecutively reads out both polarizations of Frequency 1 detectors in each pixel, moving left to right through even rows and right to left through odd rows. After all Frequency 1 detectors in a given row have been read out, consecutively read out both polarizations of Frequency 2 detectors in each pixel, moving left to right for even rows, and moving right to left for odd rows. \textit{Scheme 7} consecutively reads out both Frequency 1 polarizations then both Frequency 2 polarizations in a single pixel before reading out detectors on the next pixel, reading each row left to right. \textit{Scheme 8} consecutively reads out both Frequency 1 polarizations then both Frequency 2 polarizations in a single pixel before reading out detectors on the next pixel, reading even rows left to right and odd rows right to left.}
    \label{fig:dual_schemes}
\end{figure}

Here we use the MMT framework to evaluate the $T$-to-$B$ leakage due to crosstalk for eight different TDM readout schemes.   
These readout schemes can be separated into two categories: those that read out all detectors of one frequency before moving on to the next, which we call single-frequency, and those that interleave them, which we call dual-frequency.  All cases for each category are summarized in Figures \ref{fig:single_schemes} and \ref{fig:dual_schemes}, respectively.

\begin{figure}[t]
\centering
\includegraphics[width=1.01\linewidth]
{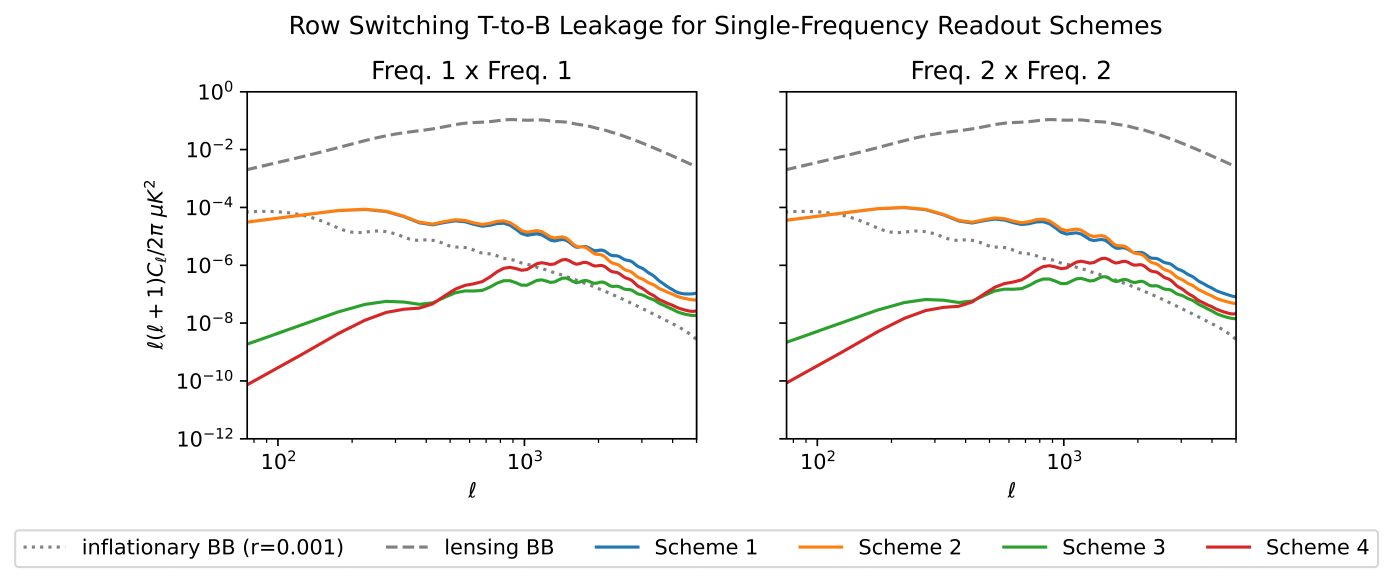}
\caption{$T$-to-$B$ leakage due to row-switching crosstalk for single-frequency readout schemes assuming crosstalk levels of 0.1\% are represented by solid lines, labeled according to Figures \ref{fig:single_schemes}. The theoretical BB spectrum for r=0.001 is indicated by the gray dotted line and the theoretical lensed BB spectrum is indicated by the gray dashed line. Each of the four panels show the leakage spectra between detectors sensitive to difference frequencies.}
\label{fig:set_time_single}
\end{figure}

\begin{figure}[t]
\centering
\includegraphics[width=1.01\linewidth]
{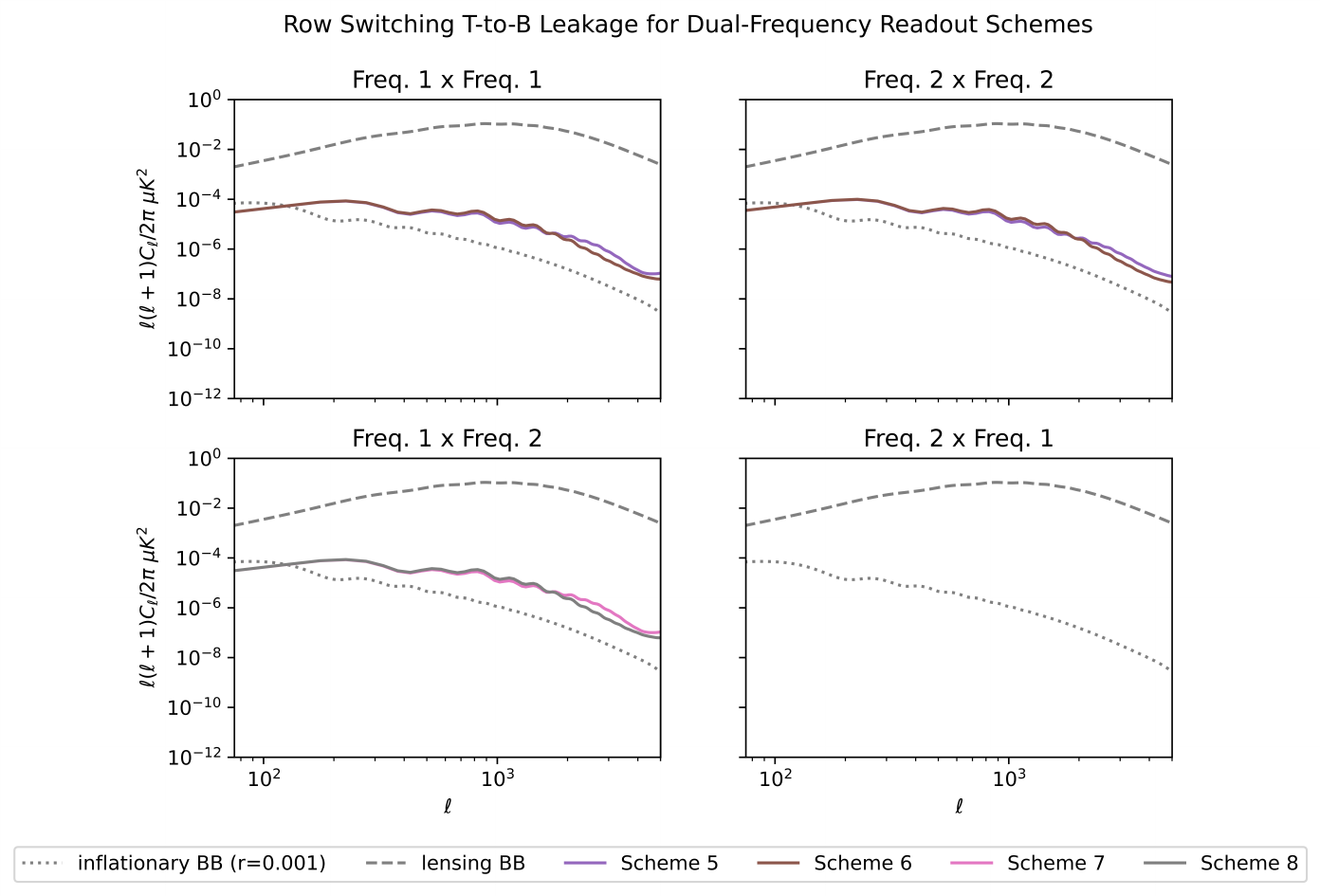}
\caption{$T$-to-$B$ leakage due to row-switching crosstalk for dual-frequency readout schemes, assuming crosstalk levels of 0.1\% are represented by solid lines, labeled according to Figures \ref{fig:dual_schemes}. The theoretical BB spectrum for r=0.001 is indicated by the gray dotted line and the theoretical lensed BB spectrum is indicated by the gray dashed line. Each of the four panels show the leakage spectra between detectors sensitive to difference frequencies.}
\label{fig:set_time_dual}
\end{figure}

\begin{figure}[t]
\centering
\includegraphics[width=1.01\linewidth]
{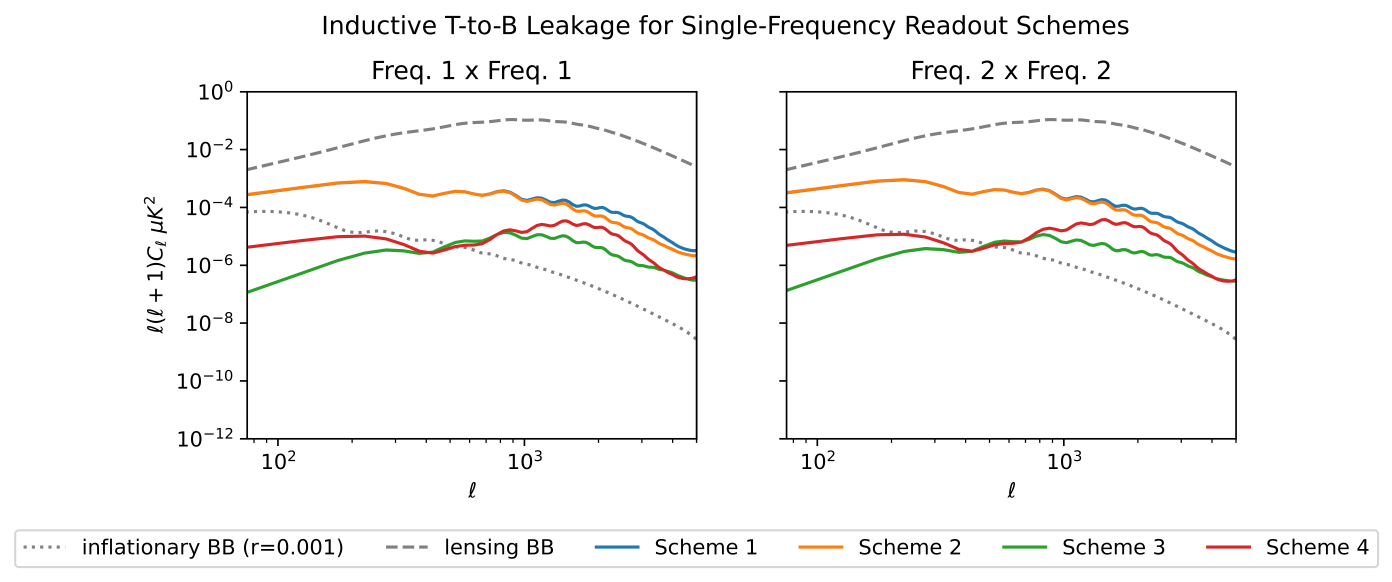}
\caption{$T$-to-$B$ leakage due to inductive crosstalk for single-frequency readout schemes, assuming crosstalk levels of 0.3\% } are represented by solid lines, labeled according to Figures \ref{fig:single_schemes}. The theoretical BB spectrum for r=0.001 is indicated by the gray dotted line and the theoretical lensed BB spectrum is indicated by the gray dashed line. Each of the four panels show the leakage spectra between detectors sensitive to difference frequencies.
\label{fig:inductive_single}
\end{figure}

\begin{figure}[t]
\centering
\includegraphics[width=1.01\linewidth]
{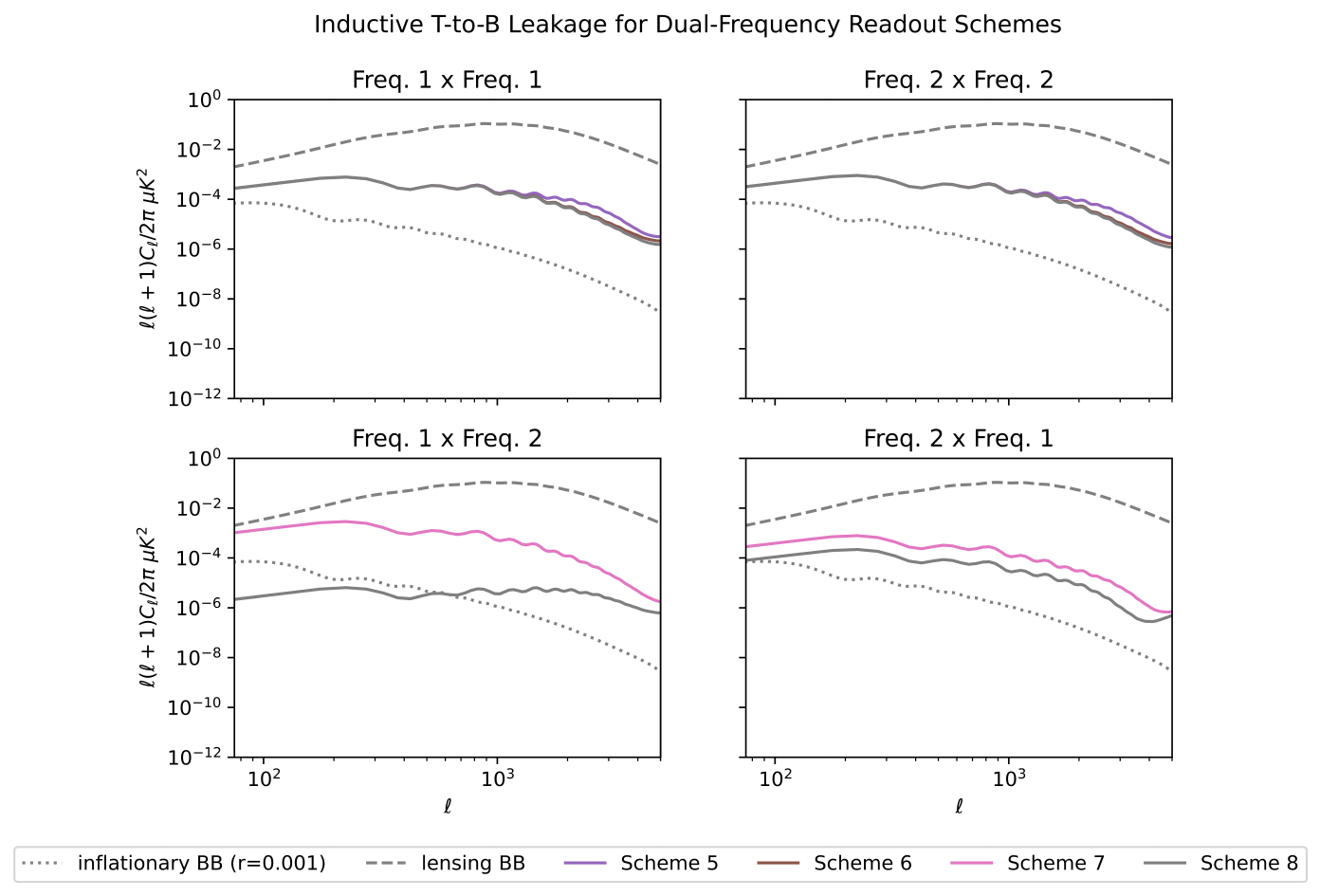}
\caption{$T$-to-$B$ leakage due to inductive crosstalk for dual-frequency readout schemes, assuming crosstalk levels of 0.3\%} are represented by solid lines, labeled according to Figures \ref{fig:dual_schemes}. The theoretical BB spectrum for r=0.001 is indicated by the gray dotted line and the theoretical dashed BB spectrum is indicated by the gray dotted line. Each of the four panels show the leakage spectra between detectors sensitive to difference frequencies.
\label{fig:inductive_dual}
\end{figure}

The resulting $T$-to-$B$ leakage signals due to row-switching crosstalk for the single-frequency and dual-frequency readout schemes are shown in Figures \ref{fig:set_time_single} and \ref{fig:set_time_dual}, while the analogous inductive crosstalk results are shown in Figures \ref{fig:inductive_single} and \ref{fig:inductive_dual}. Two theoretical B-mode signal lines are included in each plot for reference.  The lower dotted curve shows an $r = 0.001$ primordial B-mode signal, which isthe target for a CMB-S4-like experiment \cite{S4_ref_design}. The upper dashed curve is the B-mode signal caused by gravitational lensing of the E-modes; this gives a rough estimate of the point at which such systematics might become a problem for instruments deriving cosmological information from such lensing signals, or for instruments seeking to remove the lensing signal to allow measurement of the underlying primordial B-mode signal.

The shape of the leakage spectrum for each readout scheme is a consequence of the character of the corresponding $I \rightarrow (Q,U)$ leakage beams. $I \rightarrow (Q,U)$ leakage beams with monopole structure have fractional $T$-to-$B$ leakage spectra that peak at $\ell=1$, while the spectra for beams with offsetting positive and negative lobes peak at higher multipoles and decrease towards low $\ell$. We note that the details of these leakage spectra depend on the beam size and plate scale of the telescope. In this example, we consider a 5-meter, f/2.8 telescope with 5.3~mm pixel spacing.

For the single-frequency readout schemes, crosstalk occurs only between detectors operating at the same frequency, and the level of $T$-to-$B$ leakage is identical for both frequencies. For row-switching crosstalk, all schemes produce levels of leakage many orders of magnitude lower than the lensing $B$-mode spectrum, making them all suitable for experiments targeting the lensing signals. The two single-frequency schemes that read out a row of pixels in two passes (3 and 4) generate less leakage than the single-pass readout schemes (1 and 2). In particular, the leakage spectra of schemes 3 and 4 steeply decrease at low-$\ell$, while those of schemes 1 and 2 do not, making schemes 3 and 4 better suited for instruments targeting inflationary $B$-modes. For inductive crosstalk, the $T$-to-$B$ leakage spectra look similar, but with overall higher amplitudes due to the fact that the level of detector coupling is 3 times higher in the inductive case in this example.

In addition to having crosstalk between detectors operating at the same frequency, the dual-frequency readout schemes also have crosstalk between detectors of opposite frequencies frequency. For row-switching crosstalk, the schemes that read out each row in two passes (5 and 6) have crosstalk only between detectors of the same frequency since consecutively read out detectors located in different pixels are of the same frequency. The other two dual-frequency readout schemes (7 and 8), which read out each row in one pass, have crosstalk between detectors of opposite frequencies since consecutively read-out detectors located in different pixels are of different frequencies. For each pixel, the first detector read out is always the first frequency and the last detector read out is always the second frequency. Since row-switching crosstalk propagates from each detector to the next in sequence, this cross-frequency leakage is directional (Frequency 1 to Frequency 2 only). For inductive crosstalk, the $T$-to-$B$ leakage spectra look similar, but with overall higher amplitudes due to the fact that the level of detector coupling is 3 times higher in the inductive case. Additionally, the two dual-frequency schemes (7 and 8) that read out each row in one pass have crosstalk between detectors of different frequencies. Because inductive coupling links each detector to its neighbors on both sides, cross-frequency leakage occurs in both directions (Freq.~1 $\times$ Freq.~2 and Freq.~2 $\times$ Freq.~1). 

These results illustrate how MMT can efficiently quantify leakage patterns arising from readout-induced crosstalk, enabling optimization of readout architectures during instrument design. This approach can be used to model readout schemes beyond the ones presented here, including those multiplexed by other methods, and can easily be applied to larger detector arrays.

\clearpage

\subsection{Detector Time Constant Uncertainty}

CMB detectors have finite temporal response times, which must be taken account during instrument design and data analysis.  As a telescope scans across the sky, the response time effectively smears the beam along the scan direction \cite{hanany_time_const}. Although the timestream data can be deconvolved with a measured response function to correct for this effect, uncertainty in the measured time constants can leave residual response signatures in the deconvolved data, leading to a partially smeared effective beam.  It is therefore important to set requirements on the necessary precision for time constant measurements, in order
to ensure unbiased science results.

\begin{figure}[b]
\centering
\includegraphics[width=1\linewidth]{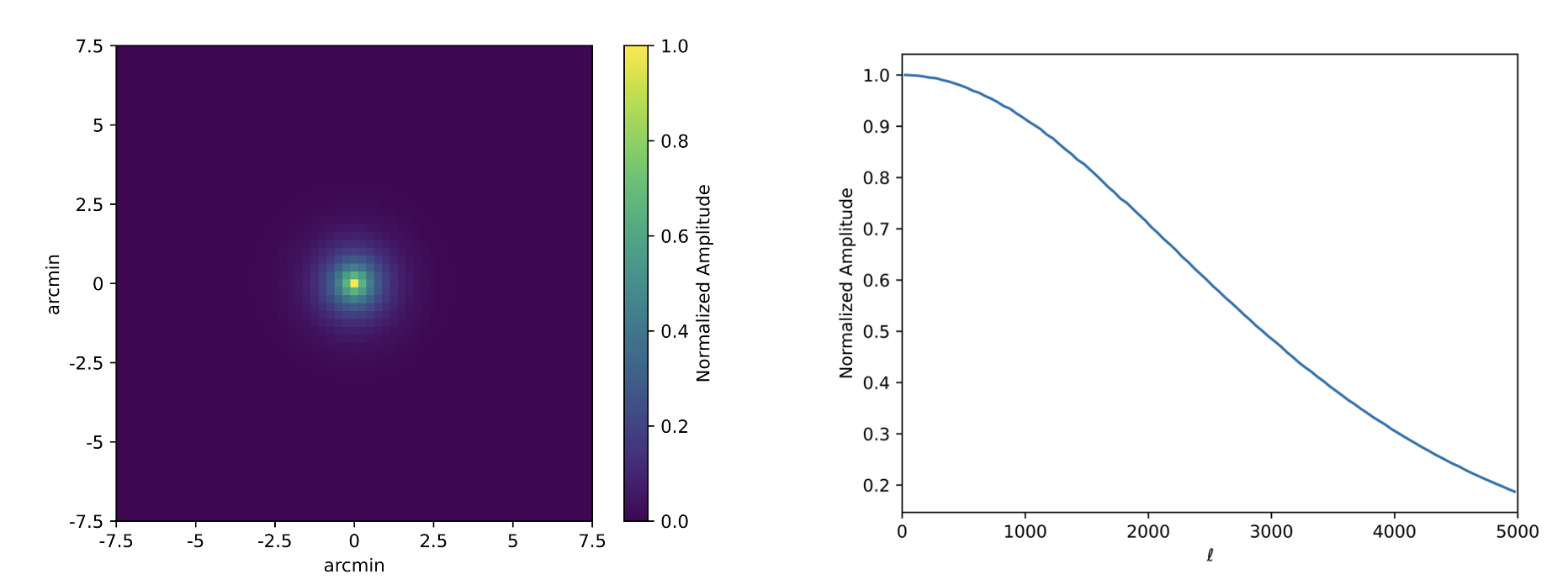}
\caption{Two-dimensional map (left panel) and one-dimensional power spectrum (right panel) of the detector response function (Equations \ref{eqn:transfer_func} and \ref{eqn:transfer_func2}) for a 10~ms detector time constant and $1^\circ$/sec scan speed assuming a perfectly cross-linked scan strategy.}
\label{fig:time_const_form}
\end{figure}

The one-dimensional detector response function $R$ can be modeled for a single detector with unity gain responding to a point source as an exponential decay
\begin{equation}
    R=e^{-\frac{\phi}{v\tau}},
    \label{eqn:transfer_func}
\end{equation}
where $v$ is the scan speed, $\phi$ is the detector's angular position on the sky, and $\tau$ is the detector time constant. In the MMT formalism, this detector response function contributes to the diagonal components of the mixing matrix $M$. However, it is illustrative to examine these components in multipole space by taking the Fourier transform of Equation \ref{eqn:transfer_func}:

\begin{equation}
    \tilde{R}=\frac{1}{\sqrt{1+{\frac{\ell^2}{\ell_{\mathit{cutoff}}^2}}}},
    \label{eqn:transfer_func2}
\end{equation}
which is the functional form of a low-pass filter with a cutoff $\ell_{\mathit{cutoff}}=180/v\tau$, as illustrated in Figure \ref{fig:time_const_form}.

Non-zero time constants therefore suppress power at high-$\ell$ in the measured CMB spectra. In the $\Lambda$CDM cosmological model, the high-$\ell$ damping tail of the temperature and $E$-mode spectra is especially sensitive to the effective number of relativistic species, $N_{\mathit{eff}}$. It is therefore important to quantify how uncertainty in detector time constant measurements propagates into biases on $N_{\mathit{eff}}$ and related cosmological parameters.

In this example, we use MMT to model the beam smear arising from varying levels of time constant uncertainty and propagate the resulting effects on the measured temperature spectrum through a cosmological parameter analysis. For simplicity, the simulation assumes a detector array with uniform $\tau$ scanning with a speed of of 1$^\circ$/sec. The scan strategy is assumed to have perfect cross-linking, which leads to uniform radial smearing of the instrument beam. The simulation also assumes that the instrument beam is perfectly deconvolved from the observed maps, so the components of the mixing matrix $M$ in this example contain only the effect of the composite detector response function for the instrument, 
which is simply $\tau$ in this case.
An error in the determination of $\tau$ is parameterized by multiplying $\tau$ in Equation \ref{eqn:transfer_func} by a factor of $(1-p)$, where $p$ is the fractional error in $\tau$.

\begin{figure}[t]
    \centering
    \includegraphics[width=1\linewidth]{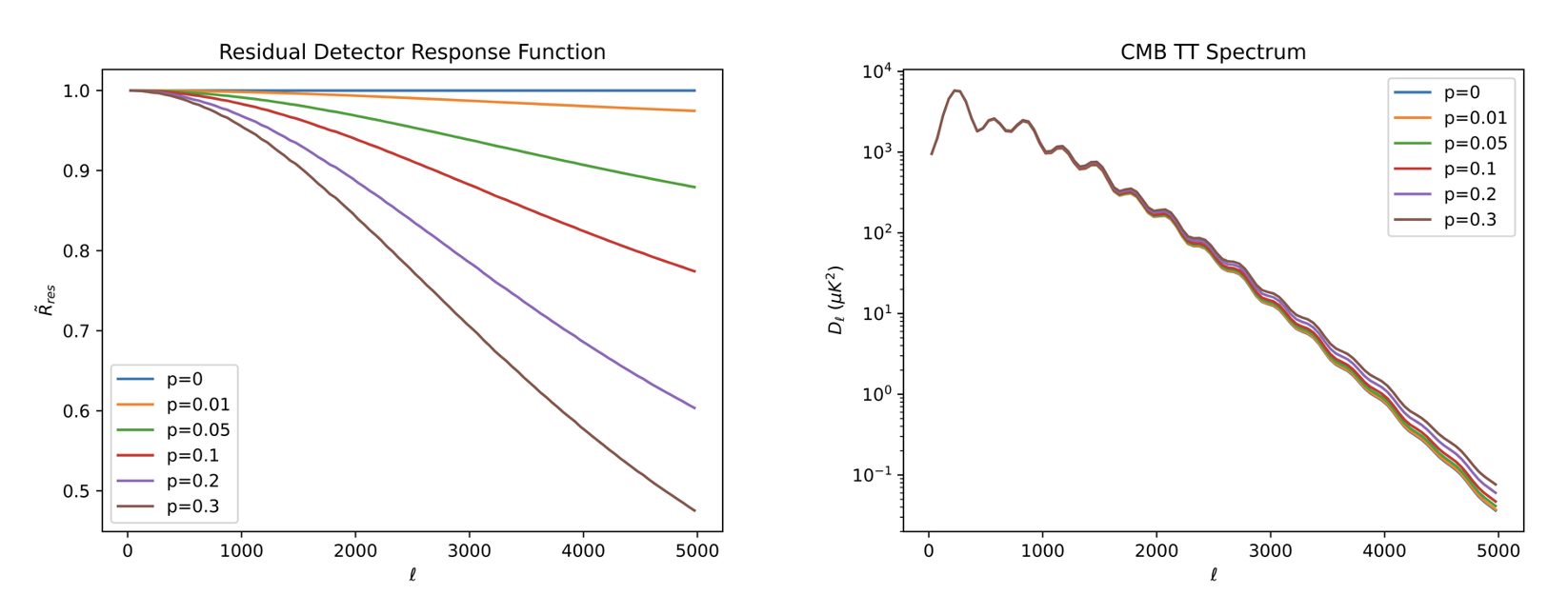}
    \caption{Left panel: Residual detector response functions $\tilde{R}_{res}$, as given by Equation \ref{eqn:residual}. Right panel: Observed CMB $TT$ spectra (after an incorrect time-constant correction) for varying levels of detector time constant uncertainty $p$ given an actual detector time constant of 10 ms with and a perfect cross-linking scan speed of $1^\circ$/sec.}
    \label{fig:time_const_effect}
\end{figure}

The residual frequency-domain response after deconvolution with the measured response function is
\begin{equation}
    \tilde{R}_{res}=\frac{\tilde{R}_{real}}{\tilde{R}_{meas}},
    \label{eqn:residual}
\end{equation}
where $\tilde{R}_{real}$ corresponds to the true response with $p=0$ and $\tilde{R}_{meas}$ represents the imperfectly measured response with nonzero $p$. $\tilde{R}_{res}$ for a range of $p$ is shown in the top panel of Figure \ref{fig:time_const_effect}. This illustrates how increasing $p$ (i.e., the measured $\tau$ being smaller than the true $\tau$) progressively lowers the effective cutoff $\ell$, suppressing the high-$\ell$ response.

Following the formalism described in Section \ref{sec:formal}, MMT generates Stokes maps of the observed CMB sky assuming a fiducial LCDM model and a fixed detector response function. These maps are then converted to power spectra, and divided by the measured detector response function for various values of $p$. The $TT$ spectra are shown in the bottom panel of Figure \ref{fig:time_const_effect}.
 
As the $p$ increases, suppression of the high-$\ell$ damping tail becomes more pronounced. These spectra are then fit with a COBAYA minimizer \cite{Cobaya} to obtain an estimate of the central values of the measured LCDM model.

\begin{figure}[t]
    \centering
    \includegraphics[width=0.5\linewidth]{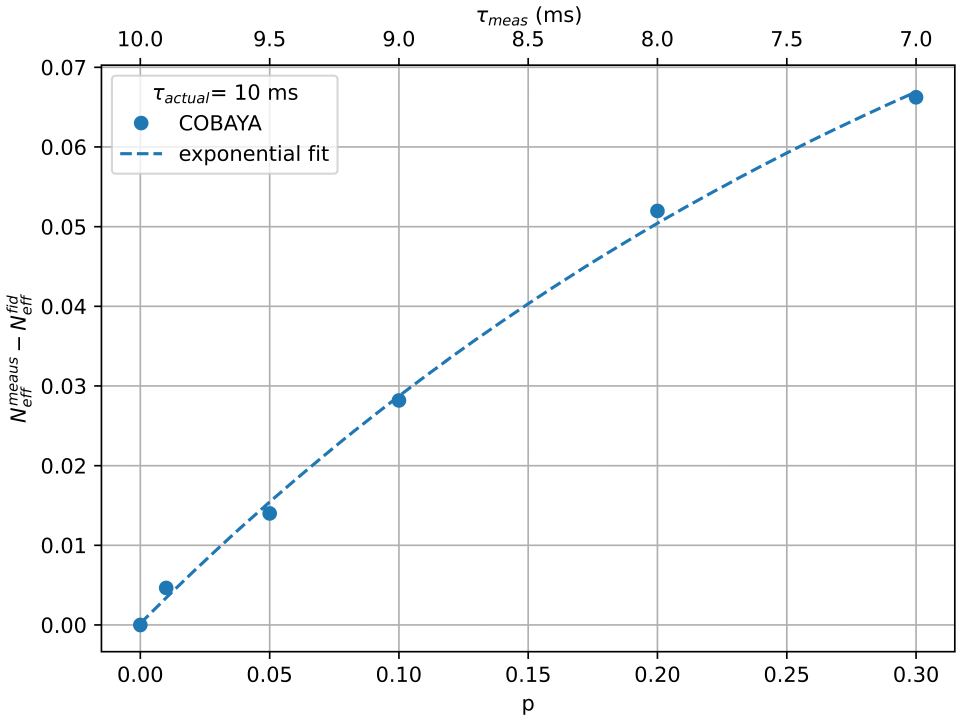}
    \caption{The effect of a range of errors $p$ in measured detector time constant $\tau_{\mathit{meas}}$ on the measured value of $N_{\mathit{eff}}$ is given by the difference of the measured ($N^{\mathit{meas}}_{\mathit{eff}}$) and fiducial ($N^{\mathit{fid}}_{\mathit{eff}}$)values. The points represent data using values of $N^{\mathit{meas}}_{\mathit{eff}}$ computed from the COBAYA minimizer, while the dashed curve shows the exponential fit. This example assumes an actual detector time constant $\tau_{\mathit{actual}}$ of 10~ms and a scan speed of $1^\circ$/sec with perfect cross-linking.}
    \label{fig:time_const_N_eff}
\end{figure}

Figure \ref{fig:time_const_N_eff} shows that as the fractional error in the measured verses true $\tau$ increases, the systematic error in determining the value of $N_{\mathit{eff}}$ increases significantly. The scale of the errors shown is non-negligible compared to the hoped-for 1$\sigma$ statistical errors of  modern experiments, such as $\sigma_{ N_{\mathit{eff}}} = 0.045$ for Simons Observatory~\cite{SO_Neff} and 
$\sigma_{N_{\mathit{eff}}} = 0.03$ for a CMB-S4-like experiment \cite{S4_ref_design}.
This example demonstrates MMT's utility in simulating systematics to determine appropriate instrument specifications to achieve an experiment's science goals.

\section{Discussion and Conclusions}
\label{sec:conclusions}
In the current era of high-sensitivity CMB experiments, MMT’s map-based framework provides an efficient and flexible approach for assessing how beam-related systematics influence observed CMB signals. MMT provides a practical means of propagating detector-level uncertainties through to impacts on cosmological parameters, enabling acceptable tolerances to be defined during experiment design. Through two representative case studies, electrical readout crosstalk and detector time constant uncertainty, we have shown that MMT can be used to efficiently quantify and compare the effects of realistic instrumental systematics. 

The electrical cross-talk example considers eight different readout schemes for a time-division multiplexed readout architecture, and demonstrates how they lead to different levels of leakage in single-frequency and multi-frequency spectra.  Depending on the target science goal, this information could be used to pick a readout scheme that minimizes the effect.  Alternatively, if measurements of the inductive and row-switching crosstalk levels are available, this could be used to understand the impact of a chosen readout scheme on the data analysis.

The detector time constant example demonstrates that these uncertainties can significantly alter high-$\ell$ CMB power spectra and therefore potentially bias derived cosmological parameters. Quantifying this relationship can be used to define calibration requirements for future instruments or check for biases in the analysis of existing datasets.

The MMT codebase and the simulation results used in this work are publicly available in \cite{PublicDataRepository}.  Although we focused on a few specific examples, the MMT framework is readily extensible to other common systematic effects, including beam ellipticity, pointing offsets, relative detector-gain variations, and far-sidelobe pickup.

\section*{Acknowledgments}
Work at CWRU was supported by DOE HEP award DE-SC0009946 and NSF award 2240374. JMN acknowledges support from RCSA award CS-CSA-2024-005. We thank Jeff Filippini and the CMB-S4 collaboration for useful discussions.

\clearpage
\bibliography{references} 

@ARTICLE{DRM_SPIE,
       author = {{Barron}, D.~R. and {Ahmed}, Z. and {Aguilar}, J. and {Anderson}, A.~J. and {Baker}, C.~F. and {Barry}, P.~S. and {Beall}, J.~A. and {Bender}, A.~N. and {Benson}, B.~A. and {Besuner}, R.~W. and {Cecil}, T.~W. and {Chang}, C.~L. and {Chapman}, S.~C. and {Chesmore}, G.~E. and {Derylo}, G. and {Doriese}, W.~B. and {Duff}, S.~M. and {Elleflot}, T. and {Filippini}, J.~P. and {Flaugher}, B. and {Gomez}, J.~G. and {Grimes}, P.~K. and {Gualtieri}, R. and {Gullett}, I. and {Haller}, G. and {Henderson}, S.~W. and {Henke}, D. and {Herbst}, R. and {Huber}, A.~I. and {Hubmayr}, J. and {Jonas}, M. and {Joseph}, J. and {King}, C.~L. and {Kovac}, J.~M. and {Kubik}, D. and {Lisovenko}, M. and {McMahon}, J.~J. and {Moncelsi}, L. and {Nagy}, J.~M. and {Osherson}, B. and {Reese}, B. and {Ruhl}, J.~E. and {Sapozhnikov}, L. and {Schillaci}, A. and {Simon}, S.~M. and {Suzuki}, A. and {Wang}, G. and {Westbrook}, B. and {Yefremenko}, V. and {Zhang}, J.},
        title = "{Conceptual design of the modular detector and readout system for the CMB-S4 survey experiment}",
    journal = {Millimeter, Submillimeter, and Far-Infrared Detectors and Instrumentation for Astronomy XI},
         year = 2022,
       editor = {{Zmuidzinas}, Jonas and {Gao}, Jian-Rong},
       series = {Society of Photo-Optical Instrumentation Engineers (SPIE) Conference Series},
       volume = {12190},
        month = aug,
          eid = {121900B},
        pages = {121900B},
          doi = {10.1117/12.2630494},
archivePrefix = {arXiv},
       eprint = {2208.02284},
 primaryClass = {astro-ph.IM},
       adsurl = {https://ui.adsabs.harvard.edu/abs/2022SPIE12190E..0BB}
}

@ARTICLE{S4_ref_design,
       author = {{Abazajian}, Kevork and {Addison}, Graeme and {Adshead}, Peter and {Ahmed}, Zeeshan and {Allen}, Steven W. and {Alonso}, David and {Alvarez}, Marcelo and {Anderson}, Adam and {Arnold}, Kam S. and {Baccigalupi}, Carlo and {Bailey}, Kathy and {Barkats}, Denis and {Barron}, Darcy and {Barry}, Peter S. and {Bartlett}, James G. and {Basu Thakur}, Ritoban and {Battaglia}, Nicholas and {Baxter}, Eric and {Bean}, Rachel and {Bebek}, Chris and {Bender}, Amy N. and {Benson}, Bradford A. and {Berger}, Edo and {Bhimani}, Sanah and {Bischoff}, Colin A. and {Bleem}, Lindsey and {Bocquet}, Sebastian and {Boddy}, Kimberly and {Bonato}, Matteo and {Bond}, J. Richard and {Borrill}, Julian and {Bouchet}, Fran{\c{c}}ois R. and {Brown}, Michael L. and {Bryan}, Sean and {Burkhart}, Blakesley and {Buza}, Victor and {Byrum}, Karen and {Calabrese}, Erminia and {Calafut}, Victoria and {Caldwell}, Robert and {Carlstrom}, John E. and {Carron}, Julien and {Cecil}, Thomas and {Challinor}, Anthony and {Chang}, Clarence L. and {Chinone}, Yuji and {Cho}, Hsiao-Mei Sherry and {Cooray}, Asantha and {Crawford}, Thomas M. and {Crites}, Abigail and {Cukierman}, Ari and {Cyr-Racine}, Francis-Yan and {de Haan}, Tijmen and {de Zotti}, Gianfranco and {Delabrouille}, Jacques and {Demarteau}, Marcel and {Devlin}, Mark and {Di Valentino}, Eleonora and {Dobbs}, Matt and {Duff}, Shannon and {Duivenvoorden}, Adriaan and {Dvorkin}, Cora and {Edwards}, William and {Eimer}, Joseph and {Errard}, Josquin and {Essinger-Hileman}, Thomas and {Fabbian}, Giulio and {Feng}, Chang and {Ferraro}, Simone and {Filippini}, Jeffrey P. and {Flauger}, Raphael and {Flaugher}, Brenna and {Fraisse}, Aurelien A. and {Frolov}, Andrei and {Galitzki}, Nicholas and {Galli}, Silvia and {Ganga}, Ken and {Gerbino}, Martina and {Gilchriese}, Murdock and {Gluscevic}, Vera and {Green}, Daniel and {Grin}, Daniel and {Grohs}, Evan and {Gualtieri}, Riccardo and {Guarino}, Victor and {Gudmundsson}, Jon E. and {Habib}, Salman and {Haller}, Gunther and {Halpern}, Mark and {Halverson}, Nils W. and {Hanany}, Shaul and {Harrington}, Kathleen and {Hasegawa}, Masaya and {Hasselfield}, Matthew and {Hazumi}, Masashi and {Heitmann}, Katrin and {Henderson}, Shawn and {Henning}, Jason W. and {Hill}, J. Colin and {Hlozek}, Ren{\'e}e and {Holder}, Gil and {Holzapfel}, William and {Hubmayr}, Johannes and {Huffenberger}, Kevin M. and {Huffer}, Michael and {Hui}, Howard and {Irwin}, Kent and {Johnson}, Bradley R. and {Johnstone}, Doug and {Jones}, William C. and {Karkare}, Kirit and {Katayama}, Nobuhiko and {Kerby}, James and {Kernovsky}, Sarah and {Keskitalo}, Reijo and {Kisner}, Theodore and {Knox}, Lloyd and {Kosowsky}, Arthur and {Kovac}, John and {Kovetz}, Ely D. and {Kuhlmann}, Steve and {Kuo}, Chao-lin and {Kurita}, Nadine and {Kusaka}, Akito and {Lahteenmaki}, Anne and {Lawrence}, Charles R. and {Lee}, Adrian T. and {Lewis}, Antony and {Li}, Dale and {Linder}, Eric and {Loverde}, Marilena and {Lowitz}, Amy and {Madhavacheril}, Mathew S. and {Mantz}, Adam and {Matsuda}, Frederick and {Mauskopf}, Philip and {McMahon}, Jeff and {McQuinn}, Matthew and {Meerburg}, P. Daniel and {Melin}, Jean-Baptiste and {Meyers}, Joel and {Millea}, Marius and {Mohr}, Joseph and {Moncelsi}, Lorenzo and {Mroczkowski}, Tony and {Mukherjee}, Suvodip and {M{\"u}nchmeyer}, Moritz and {Nagai}, Daisuke and {Nagy}, Johanna and {Namikawa}, Toshiya and {Nati}, Federico and {Natoli}, Tyler and {Negrello}, Mattia and {Newburgh}, Laura and {Niemack}, Michael D. and {Nishino}, Haruki and {Nordby}, Martin and {Novosad}, Valentine and {O'Connor}, Paul and {Obied}, Georges and {Padin}, Stephen and {Pandey}, Shivam and {Partridge}, Bruce and {Pierpaoli}, Elena and {Pogosian}, Levon and {Pryke}, Clement and {Puglisi}, Giuseppe and {Racine}, Benjamin and {Raghunathan}, Srinivasan and {Rahlin}, Alexandra and {Rajagopalan}, Srini and {Raveri}, Marco and {Reichanadter}, Mark and {Reichardt}, Christian L. and {Remazeilles}, Mathieu and {Rocha}, Graca and {Roe}, Natalie A. and {Roy}, Anirban and {Ruhl}, John and {Salatino}, Maria and {Saliwanchik}, Benjamin and {Schaan}, Emmanuel and {Schillaci}, Alessandro and {Schmittfull}, Marcel M. and {Scott}, Douglas and {Sehgal}, Neelima and {Shandera}, Sarah and {Sheehy}, Christopher and {Sherwin}, Blake D. and {Shirokoff}, Erik and {Simon}, Sara M. and {Slosar}, Anze and {Somerville}, Rachel and {Spergel}, David and {Staggs}, Suzanne T. and {Stark}, Antony and {Stompor}, Radek and {Story}, Kyle T. and {Stoughton}, Chris and {Suzuki}, Aritoki and {Tajima}, Osamu and {Teply}, Grant P. and {Thompson}, Keith and {Timbie}, Peter and {Tomasi}, Maurizio and {Treu}, Jesse I. and {Tristram}, Matthieu and {Tucker}, Gregory and {Umilt{\`a}}, Caterina and {van Engelen}, Alexander and {Vieira}, Joaquin D. and {Vieregg}, Abigail G. and {Vogelsberger}, Mark and {Wang}, Gensheng and {Watson}, Scott and {White}, Martin and {Whitehorn}, Nathan and {Wollack}, Edward J. and {Kimmy Wu}, W.~L. and {Xu}, Zhilei and {Yasini}, Siavash and {Yeck}, James and {Yoon}, Ki Won and {Young}, Edward and {Zonca}, Andrea},
        title = "{CMB-S4 Science Case, Reference Design, and Project Plan}",
      journal = {arXiv e-prints},
         year = 2019,
        month = jul,
          eid = {arXiv:1907.04473},
        pages = {arXiv:1907.04473},
          doi = {10.48550/arXiv.1907.04473},
archivePrefix = {arXiv},
       eprint = {1907.04473},
 primaryClass = {astro-ph.IM},
}

@ARTICLE{S4_science_book,
       author = {{Abazajian}, Kevork N. and {Adshead}, Peter and {Ahmed}, Zeeshan and
         {Allen}, Steven W. and {Alonso}, David and {Arnold}, Kam S. and
         {Baccigalupi}, Carlo and {Bartlett}, James G. and
         {Battaglia}, Nicholas and {Benson}, Bradford A. and
         {Bischoff}, Colin A. and {Borrill}, Julian and {Buza}, Victor and
         {Calabrese}, Erminia and {Caldwell}, Robert and {Carlstrom}, John E. and
         {Chang}, Clarence L. and {Crawford}, Thomas M. and
         {Cyr-Racine}, Francis-Yan and {De Bernardis}, Francesco and
         {de Haan}, Tijmen and {di Serego Alighieri}, Sperello and
         {Dunkley}, Joanna and {Dvorkin}, Cora and {Errard}, Josquin and
         {Fabbian}, Giulio and {Feeney}, Stephen and {Ferraro}, Simone and
         {Filippini}, Jeffrey P. and {Flauger}, Raphael and {Fuller}, George M. and
         {Gluscevic}, Vera and {Green}, Daniel and {Grin}, Daniel and
         {Grohs}, Evan and {Henning}, Jason W. and {Hill}, J. Colin and
         {Hlozek}, Renee and {Holder}, Gilbert and {Holzapfel}, William and
         {Hu}, Wayne and {Huffenberger}, Kevin M. and {Keskitalo}, Reijo and
         {Knox}, Lloyd and {Kosowsky}, Arthur and {Kovac}, John and
         {Kovetz}, Ely D. and {Kuo}, Chao-Lin and {Kusaka}, Akito and
         {Le Jeune}, Maude and {Lee}, Adrian T. and {Lilley}, Marc and
         {Loverde}, Marilena and {Madhavacheril}, Mathew S. and {Mantz}, Adam and
         {Marsh}, David J.~E. and {McMahon}, Jeffrey and
         {Meerburg}, Pieter Daniel and {Meyers}, Joel and {Miller}, Amber D. and
         {Munoz}, Julian B. and {Nguyen}, Ho Nam and {Niemack}, Michael D. and
         {Peloso}, Marco and {Peloton}, Julien and {Pogosian}, Levon and
         {Pryke}, Clement and {Raveri}, Marco and {Reichardt}, Christian L. and
         {Rocha}, Graca and {Rotti}, Aditya and {Schaan}, Emmanuel and
         {Schmittfull}, Marcel M. and {Scott}, Douglas and {Sehgal}, Neelima and
         {Shandera}, Sarah and {Sherwin}, Blake D. and {Smith}, Tristan L. and
         {Sorbo}, Lorenzo and {Starkman}, Glenn D. and {Story}, Kyle T. and
         {van Engelen}, Alexander and {Vieira}, Joaquin D. and {Watson}, Scott and
         {Whitehorn}, Nathan and {Kimmy Wu}, W.~L.},
        title = "{CMB-S4 Science Book, First Edition}",
      journal = {arXiv e-prints},
         year = "2016",
        month = "Oct",
          eid = {arXiv:1610.02743},
        pages = {arXiv:1610.02743},
archivePrefix = {arXiv},
       eprint = {1610.02743},
 primaryClass = {astro-ph.CO},
}

@article{BKS_dets,
  title={Antenna-coupled TES bolometers used in BICEP2, Keck array, and SPIDER},
  author={Ade, Peter AR and Aikin, RW and Amiri, M and Barkats, D and Benton, SJ and Bischoff, Colin A and Bock, JJ and Bonetti, JA and Brevik, JA and Buder, I and others},
  journal={The Astrophysical Journal},
  volume={812},
  number={2},
  pages={176},
  year={2015},
  publisher={IOP Publishing}
}

@article{spt3g,
  title={The design and integrated performance of SPT-3G},
  author={Sobrin, JA and Anderson, AJ and Bender, AN and Benson, BA and Dutcher, D and Foster, A and Goeckner-Wald, N and Montgomery, J and Nadolski, A and Rahlin, A and others},
  journal={The Astrophysical Journal Supplement Series},
  volume={258},
  number={2},
  pages={42},
  year={2022},
  publisher={IOP Publishing}
}

@article{BK18_2021,
	adsurl = {https://ui.adsabs.harvard.edu/abs/2021PhRvL.127o1301A},
	archiveprefix = {arXiv},
	author = {{BICEP/Keck Collaboration} and {Ade}, P.~A.~R. and {Ahmed}, Z. and {Amiri}, M. and {Barkats}, D. and {Thakur}, R. Basu and {Bischoff}, C.~A. and {Beck}, D. and {Bock}, J.~J. and {Boenish}, H. and {Bullock}, E. and {Buza}, V. and {Cheshire}, J.~R. and {Connors}, J. and {Cornelison}, J. and {Crumrine}, M. and {Cukierman}, A. and {Denison}, E.~V. and {Dierickx}, M. and {Duband}, L. and {Eiben}, M. and {Fatigoni}, S. and {Filippini}, J.~P. and {Fliescher}, S. and {Goeckner-Wald}, N. and {Goldfinger}, D.~C. and {Grayson}, J. and {Grimes}, P. and {Hall}, G. and {Halal}, G. and {Halpern}, M. and {Hand}, E. and {Harrison}, S. and {Henderson}, S. and {Hildebrandt}, S.~R. and {Hilton}, G.~C. and {Hubmayr}, J. and {Hui}, H. and {Irwin}, K.~D. and {Kang}, J. and {Karkare}, K.~S. and {Karpel}, E. and {Kefeli}, S. and {Kernasovskiy}, S.~A. and {Kovac}, J.~M. and {Kuo}, C.~L. and {Lau}, K. and {Leitch}, E.~M. and {Lennox}, A. and {Megerian}, K.~G. and {Minutolo}, L. and {Moncelsi}, L. and {Nakato}, Y. and {Namikawa}, T. and {Nguyen}, H.~T. and {O'Brient}, R. and {Ogburn}, R.~W. and {Palladino}, S. and {Prouve}, T. and {Pryke}, C. and {Racine}, B. and {Reintsema}, C.~D. and {Richter}, S. and {Schillaci}, A. and {Schwarz}, R. and {Schmitt}, B.~L. and {Sheehy}, C.~D. and {Soliman}, A. and {Germaine}, T. St. and {Steinbach}, B. and {Sudiwala}, R.~V. and {Teply}, G.~P. and {Thompson}, K.~L. and {Tolan}, J.~E. and {Tucker}, C. and {Turner}, A.~D. and {Umilt{\`a}}, C. and {Verg{\`e}s}, C. and {Vieregg}, A.~G. and {Wandui}, A. and {Weber}, A.~C. and {Wiebe}, D.~V. and {Willmert}, J. and {Wong}, C.~L. and {Wu}, W.~L.~K. and {Yang}, H. and {Yoon}, K.~W. and {Young}, E. and {Yu}, C. and {Zeng}, L. and {Zhang}, C. and {Zhang}, S. and {Bicep/Keck Collaboration}},
	doi = {10.1103/PhysRevLett.127.151301},
	eid = {151301},
	eprint = {2110.00483},
	journal = {\prl},
	month = oct,
	number = {15},
	pages = {151301},
	primaryclass = {astro-ph.CO},
	title = {{Improved Constraints on Primordial Gravitational Waves using Planck, WMAP, and BICEP/Keck Observations through the 2018 Observing Season}},
	volume = {127},
	year = 2021}

@article{horn_dets,
  title={Horn coupled multichroic polarimeters for the Atacama Cosmology Telescope polarization experiment},
  author={Datta, Rahul and Hubmayr, Johannes and Munson, Charles and Austermann, Jason and Beall, James and Becker, Dan and Cho, Hsiao-Mei and Halverson, Nils and Hilton, Gene and Irwin, Kent and others},
  journal={Journal of Low Temperature Physics},
  volume={176},
  pages={670--676},
  year={2014},
  publisher={Springer}
}

@article{act,
  title={The Atacama cosmology telescope: the polarization-sensitive ACTPol instrument},
  author={Thornton, RJ and Ade, PAR and Aiola, S and Angile, FE and Amiri, M and Beall, JA and Becker, DT and Cho, HM and Choi, SK and Corlies, P and others},
  journal={The Astrophysical Journal Supplement Series},
  volume={227},
  number={2},
  pages={21},
  year={2016},
  publisher={IOP Publishing}
}

@inproceedings{SO,
  title={The Simons observatory: instrument overview},
  author={Galitzki, Nicholas and Ali, Aamir and Arnold, Kam S and Ashton, Peter C and Austermann, Jason E and Baccigalupi, Carlo and Baildon, Taylor and Barron, Darcy and Beall, James A and Beckman, Shawn and others},
  booktitle={Millimeter, Submillimeter, and Far-Infrared Detectors and Instrumentation for Astronomy IX},
  volume={10708},
  pages={40--52},
  year={2018},
  organization={SPIE}
}

@article{Hu_Hedman_Zaldarriaga,
  title = {Benchmark parameters for CMB polarization experiments},
  author = {Hu, Wayne and Hedman, Matthew M. and Zaldarriaga, Matias},
  journal = {Phys. Rev. D},
  volume = {67},
  issue = {4},
  pages = {043004},
  numpages = {11},
  year = {2003},
  month = {Feb},
  publisher = {American Physical Society},
  doi = {10.1103/PhysRevD.67.043004},
  url = {https://link.aps.org/doi/10.1103/PhysRevD.67.043004}
}

@article{shimon_keating,
  title={CMB polarization systematics due to beam asymmetry: Impact on inflationary science},
  author={Shimon, Meir and Keating, Brian and Ponthieu, Nicolas and Hivon, Eric},
  journal={Physical Review D—Particles, Fields, Gravitation, and Cosmology},
  volume={77},
  number={8},
  pages={083003},
  year={2008},
  publisher={APS}
}

@article{Planck_tod_sims,
  title={A simulation pipeline for the Planck mission},
  author={Reinecke, Martin and Dolag, Klaus and Hell, Reinhard and Bartelmann, Matthias and En{\ss}lin, TA},
  journal={Astronomy \& Astrophysics},
  volume={445},
  number={1},
  pages={373--373},
  year={2006},
  publisher={EDP Sciences}
}

@article{TOAST,
  title={Simulating Calibration and Beam Systematics for a Future CMB Space Mission with the TOAST Package},
  author={Puglisi, Giuseppe and Keskitalo, Reijo and Kisner, Ted and Borrill, Julian D},
  journal={Research Notes of the AAS},
  volume={5},
  number={6},
  pages={137},
  year={2021},
  publisher={The American Astronomical Society}
}

@article{Class_Vpol,
  title={Two-year cosmology large angular scale surveyor (CLASS) observations: A measurement of circular polarization at 40 GHz},
  author={Padilla, Ivan L and Eimer, Joseph R and Li, Yunyang and Addison, Graeme E and Ali, Aamir and Appel, John W and Bennett, Charles L and Bustos, Ricardo and Brewer, Michael K and Chan, Manwei and others},
  journal={The Astrophysical Journal},
  volume={889},
  number={2},
  pages={105},
  year={2020},
  publisher={IOP Publishing}
}

@article{Spider_Vpol,
  title={A new limit on CMB circular polarization from SPIDER},
  author={Nagy, JM and Ade, PAR and Amiri, M and Benton, SJ and Bergman, AS and Bihary, R and Bock, JJ and Bond, JR and Bryan, SA and Chiang, HC and others},
  journal={The Astrophysical Journal},
  volume={844},
  number={2},
  pages={151},
  year={2017},
  publisher={IOP Publishing}
}

@article{Cobaya,
  title={Cobaya: Code for Bayesian Analysis of hierarchical physical models},
  author={Torrado, Jesus and Lewis, Antony},
  journal={Journal of Cosmology and Astroparticle Physics},
  volume={2021},
  number={05},
  pages={057},
  year={2021},
  publisher={IOP Publishing}
}

@misc{PublicDataRepository,
    author = "{A. Hryciuk}",
    title = {{Public Data Repository}},
    howpublished = {\url{\texttt{https://github.com/CMB-S4/map\textunderscore multi\textunderscore tool}}},
    note = {Online; accessed May 2026} ,
    year=2025,
}

@article{SO_science_book,
  title={The Simons Observatory: science goals and forecasts},
  author={Ade, Peter and Aguirre, James and Ahmed, Zeeshan and Aiola, Simone and Ali, Aamir and Alonso, David and Alvarez, Marcelo A and Arnold, Kam and Ashton, Peter and Austermann, Jason and others},
  journal={Journal of Cosmology and Astroparticle Physics},
  volume={2019},
  number={02},
  pages={056},
  year={2019},
  publisher={IOP Publishing}
}

@article{SO_upgrade,
  title={The Simons Observatory: Science Goals and Forecasts for the Enhanced Large Aperture Telescope},
  author={Abitbol, M and Abril-Cabezas, I and Adachi, S and Ade, P and Adler, AE and Agrawal, P and Aguirre, J and Ahmed, Z and Aiola, S and Alford, T and others},
  journal={arXiv preprint arXiv:2503.00636},
  year={2025}
}

@article{bicep2_pairdiff,
  title={BICEP2. II. Experiment and three-year Data Set},
  author={Ade, Peter AR and Aikin, RW and Amiri, M and Barkats, Denis and Benton, SJ and Bischoff, CA and Bock, JJ and Brevik, JA and Buder, I and Bullock, E and others},
  journal={The Astrophysical Journal},
  volume={792},
  number={1},
  pages={62},
  year={2014},
  publisher={IOP Publishing}
}

@article{polarbear_pairdiff,
  title={Making maps of cosmic microwave background polarization for B-mode studies: the POLARBEAR example},
  author={Poletti, Davide and Fabbian, Giulio and Le Jeune, Maude and Peloton, Julien and Arnold, Kam and Baccigalupi, Carlo and Barron, Darcy and Beckman, Shawn and Borrill, Julian and Chapman, Scott and others},
  journal={Astronomy \& Astrophysics},
  volume={600},
  pages={A60},
  year={2017},
  publisher={EDP Sciences}
}

@article{S4_TDM,
  title={End-to-end modeling of the TDM readout system for CMB-S4},
  author={Goldfinger, David C and Ahmed, Zeeshan and Barron, Darcy R and Doriese, W Bertrand and Durkin, Malcolm and Filippini, Jeffrey P and Haller, Gunther and Henderson, Shawn W and Herbst, Ryan and Hubmayr, Johannes and others},
  journal={Journal of Low Temperature Physics},
  volume={215},
  number={3},
  pages={143--151},
  year={2024},
  publisher={Springer}
}

@article{TDM_2level,
  title={Symmetric time-division-multiplexed SQUID readout with two-layer switches for future TES observatories},
  author={Durkin, Malcolm and Backhaus, Scott and Bandler, Simon R and Chervenak, James A and Denison, Ed V and Doriese, William B and Gard, Johnathon D and Hilton, Gene C and Lew, Richard A and Lucas, Tammy J and others},
  journal={IEEE Transactions on Applied Superconductivity},
  volume={33},
  number={5},
  pages={1--5},
  year={2023},
  publisher={IEEE}
}

@article{EB_conversion,
  title={Nature of the E-B decomposition of CMB polarization},
  author={Zaldarriaga, Matias},
  journal={Physical Review D},
  volume={64},
  number={10},
  pages={103001},
  year={2001},
  publisher={APS}
}

@article{hanany_time_const,
  title={The effect of the detector response time on bolometric cosmic microwave background anisotropy experiments},
  author={Hanany, S and Jaffe, Andrew H and Scannapieco, E},
  journal={Monthly Notices of the Royal Astronomical Society},
  volume={299},
  number={3},
  pages={653--660},
  year={1998},
  publisher={The Royal Astronomical Society}
}

@article{padin2018three,
  title={Three-mirror anastigmat for cosmic microwave background observations},
  author={Padin, S},
  journal={Applied Optics},
  volume={57},
  number={9},
  pages={2314--2326},
  year={2018},
  publisher={Optical Society of America}
}

@article{Gallardo24,
author = {Patricio A. Gallardo and Roberto Puddu and Kathleen Harrington and Bradford Benson and John E. Carlstrom and Simon R. Dicker and Nick Emerson and Jon E. Gudmundsson and Michele Limon and Jeff McMahon and Johanna M. Nagy and Tyler Natoli and Michael D. Niemack and Stephen Padin and John Ruhl and Sara M. Simon and The CMB-S4 Collaboration},
journal = {Appl. Opt.},
number = {2},
pages = {310--321},
publisher = {Optica Publishing Group},
title = {Freeform three-mirror anastigmatic large-aperture telescope andreceiver optics for CMB-S4},
volume = {63},
month = {Jan},
year = {2024},
url = {https://opg.optica.org/ao/abstract.cfm?URI=ao-63-2-310},
doi = {10.1364/AO.501744},

}

@article{SO_Neff,
  title={The Simons Observatory: science goals and forecasts for the enhanced Large Aperture Telescope},
  author={Abitbol, M and Abril-Cabezas, I and Adachi, S and Ade, P and Adler, AE and Agrawal, P and Aguirre, J and Ahmed, Z and Aiola, S and Alford, T and others},
  journal={Journal of Cosmology and Astroparticle Physics},
  volume={2025},
  number={08},
  pages={034},
  year={2025},
  publisher={IOP Publishing}
  }

@article{bicep2_deproj,
  title={Bicep2. III. Instrumental Systematics},
  author={Ade, PAR and Aikin, RW and Barkats, Denis and Benton, SJ and Bischoff, CA and Bock, JJ and Brevik, JA and Buder, I and Bullock, E and Dowell, CD and others},
  journal={The Astrophysical Journal},
  volume={814},
  number={2},
  pages={110},
  year={2015},
  publisher={The American Astronomical Society}
}

@article{mapmaking_better,
  title={ROMA: A map-making algorithm for polarised CMB data sets},
  author={De Gasperis, Giancarlo and Balbi, Amedeo and Cabella, Paolo and Natoli, Paolo and Vittorio, Nicola},
  journal={Astronomy \& Astrophysics},
  volume={436},
  number={3},
  pages={1159--1165},
  year={2005},
  publisher={EDP Sciences}
}

\end{document}